\documentclass[twocolumn,showpacs,prl]{revtex4}
\usepackage{epsfig}
\usepackage{graphicx}
\newcommand{\vnabla}{{\mbox{\boldmath$\nabla$}}}
\newcommand{\veta}{{\mbox{\boldmath$\eta$}}}

\newcommand{\vR}{{\mbox{\boldmath$R$}}}
\newcommand{\vD}{{\mbox{\boldmath$D$}}}

\newcommand{\vA}{{\mbox{\boldmath$A$}}}
\newcommand{\va}{{\mbox{\boldmath$a$}}}
\newcommand{\br}{{\mbox{\boldmath$r$}}}

\newcommand{\vb}{{\mbox{\boldmath$b$}}}
\newcommand{\vB}{{\mbox{\boldmath$B$}}}
\newcommand{\vtau}{{\mbox{\boldmath$\tau$}}}

\newcommand{\vq}{\mbox{\boldmath$q$}}
\newcommand{\vpsi}{\mbox{\boldmath$\psi$}}

\begin{document}

\title{Spatial line nodes and fractional vortex pairs in the Fulde-Ferrell-Larkin-Ovchinnikov phase}
\author{D.F. Agterberg, Z. Zheng, and S. Mukherjee}
\address{Department of Physics, University of Wisconsin-Milwaukee, Milwaukee, WI 53211}
%\PACS{74.20.De,74.25.Dw }

\begin{abstract}
A Zeeman magnetic field can induce a
Fulde-Ferrell-Larkin-Ovchinnikov (FFLO) phase in spin-singlet
superconductors.  Here we argue that there is a non-trivial
solution for the FFLO vortex phase that exists near the upper
critical field in which the wavefunction has only spatial line
nodes that form intricate and unusual three-dimensional
structures. These structures include a crisscrossing lattice of
two sets of non-parallel line nodes. We show that these solutions
arise from the decay of conventional Abrikosov vortices into pairs
of fractional vortices. We propose that neutron scattering studies
can observe these fractional vortex pairs through the observation
of a lattice of 1/2 flux quanta vortices. We also consider related
phases in non-centrosymmetric (NC) superconductors.
\end{abstract} \maketitle

A FFLO phase predicted in Refs.~\cite{lar65,ful64} appears to have
been discovered in CeCoIn$_5$ in the high magnetic field region of
the superconducting phase diagram \cite{rad03,bia03}.  This
discovery has generated tremendous interest both experimentally
and theoretically \cite{mat07}. FFLO phases have also been argued
to be of importance in understanding ultracold atomic Fermi gases
\cite{zwi06} and in the formation of color superconductivity in
high density quark matter \cite{cas04}. The understanding of these
phases has become a relevant and topical pursuit in physics. One
central issue is the role vortices play in these phases: in
CeCoIn$_5$ the FFLO phase appears deep within a vortex phase
\cite{rad03,bia03}; and ultracold atomic Fermi gases can be
rotated to create vortices within an FFLO phase \cite{shi06}.

Here we address the nature of the FFLO vortex phase. Previous
studies have concluded that the superconducting gap function in
this phase is, for example, $\Delta(\vR)=\cos(qz)\phi_n(\br)$
where the magnetic field is applied along the $\hat{z}$ direction,
$\hat{z}\cdot\br=0$, and $\phi_n(\br)$ describes a vortex lattice
constructed from a Landau level (LL) with index $n$
\cite{tac96,hou01,yan04,miz05,hou06}. This solution has
intersecting spatial nodes along planes perpendicular to the
$z$-axis and along the vortex lines parallel to the $z$-axis. We
show that there is another realistic solution for the FFLO vortex
phase in which there are only spatial line nodes in the gap
function.  We show that the existence of this solution is a
consequence of the decay of conventional  vortices into  pairs of
fractional vortices. These fractional vortices exist because of
the broken translational symmetry inherent in FFLO
superconductors. By suitably choosing an order parameter that
correctly exhibits this broken translational symmetry, these
fractional vortices naturally appear within the theory. We propose
that a small angle neutron scattering (SANS) measurement of the
resulting magnetic field distribution may observe a lattice of 1/2
flux quanta near to the upper critical field.  We further argue
that this phase is stable within weak-coupling theories of
superconductivity and consider related phases in NC
superconductors.

We use a phenomenological approach pioneered by Buzdin and
Kachkachi to describe the FFLO phase
\cite{buz97,hou01,yan04,hou06}, and extend it to include NC
superconductors. We begin with the following free energy
\begin{eqnarray}
F=& \int d^3R \{ \alpha|\Delta|^2+\beta|\Delta|^4+\nu|\Delta|^6+\kappa|\vD\Delta|^2+\delta|\vD^2\Delta|^2\nonumber \\ &+
\mu|\Delta|^2|\vD\Delta|^2+\eta[(\Delta^*)^2(\vD \Delta)^2+
(\Delta)^2(\vD^* \Delta^*)^{2}]\nonumber \\& +\epsilon\vB\cdot[\Delta^*(\vD\Delta)+\Delta(\vD\Delta)^*] \}\label{free}
\end{eqnarray} where  $\vD=-i\vnabla- 2e \vA$ and $\vB=\nabla\times\vA$. The coefficients that appear in this free energy are typically determined from a microscopic BCS theory \cite{buz97}. The $\epsilon$-term applies only to NC superconductors.  It results in the helical phase discussed previously \cite{kau05}.  In this phase, the gap function becomes $\Delta(\vR)=\psi_1e^{i\vq\cdot\vR}$. The orientation of $\vq$ is determined by the free energy invariant denoted by $\epsilon$ in Eq.~\ref{free}. We have chosen this invariant so that the theory applies to Li$_2$Pt$_3$B with point group $O$ \cite{yua06}. Consequently, $\vq$ is parallel to $\vB$. With $\epsilon=0$, Eq.~\ref{free} has been justified previously \cite{buz97}.

We consider a magnetic field along the $\hat{z}$ direction and ignore screening currents in determining the high field ground state structure of the gap function (this is reasonable for strongly type II superconductors). In the normal state there will be translational invariance along the magnetic field direction.  Therefore Fourier modes along this direction will be eigenstates of the linear gap equation. Typically, the eigenstate with the lowest energy corresponds to the Fourier mode $q=0$. However in FFLO superconductors, the eigenstate with the lowest energy have finite $q$. The states $\pm q$ are degenerate and this degeneracy is broken by non-linear terms in the free energy. Consequently, to describe the FFLO phase near the upper critical field, it suffices to keep the two modes $\pm q$. We therefore write  $\Delta(\vR)=\psi_1(\br)e^{iqz}+\psi_2(\br)e^{-iqz}$ where $\br$ is orthogonal to the magnetic field and $\vq$ is parallel to the field. This yields the following free energy for the new order parameter $\vpsi=(\psi_1,\psi_2)$:
\begin{eqnarray}
F=&L_z\int d^2r \{ \alpha_1|\psi_1|^2+\alpha_2|\psi_2|^2+\beta_1|\vpsi|^4+\beta_2|\psi_1|^2|\psi_2|^2\nonumber \\ &\nu|\vpsi|^6+6\nu|\vpsi|^2|\psi_1|^2|\psi_2|^2\nonumber\\
&+\kappa_1(|\vD\psi_1|^2+|\vD\psi_2|^2) +\kappa_2(|\vD^2\psi_1|^2+|\vD^2\psi_2|^2)\nonumber\\
&+\mu(|\vD \psi_1|^2|\psi_1|^2+|\vD \psi_2|^2|\psi_2|^2)\nonumber \\&+\eta[(\vD\psi_1)^2(\psi_1^*)^2+(\vD\psi_2)^2(\psi_2^*)^2+c.c]\nonumber \\ & +4\eta[(\vD\psi_1)\psi_1^*(\vD\psi_2)\psi_2^*+c.c]\nonumber\\ & +\mu[(\vD\psi_1)\psi_1^*(\vD\psi_2)^*\psi_2+c.c]\nonumber\\ &
\mu[|\vD\psi_1|^2|\psi_2|^2+|\vD\psi_2|^2|\psi_1|^2]\}
\label{free1}
\end{eqnarray}
where $L_z$ is the size of the system along the $z$ direction, $\vD=(D_x,D_y)$ and $c.c.$ means complex conjugate. The coefficients in Eq.~\ref{free1} now depend upon $q$ \cite{eq2}. For FFLO superconductors $\alpha_1=\alpha_2$. Eq.~\ref{free1} should be optimized with respect to $q$ and we assume this has been done. This ensures that there is no net current flowing along the $z$ direction \cite{hou01,kau05}.

The choice of order parameter $\vpsi$ manifestly exhibits the broken translational symmetry that characterizes the FFLO state. This broken symmetry is hidden when considering $\Delta$.  By considering $\vpsi$ explicitly, new and general features of the theory appear naturally. In particular, notice that Eq.~\ref{free1} is independent of separate rotations of the phases of $\psi_1$ and $\psi_2$, revealing a global $U(1)\times U(1)$ gauge invariance. This  follows from translational invariance of the normal state along the $z$ direction and usual gauge invariance. In particular, consider a general term $\psi_1^n\psi_2^m(\psi_1^*)^p(\psi_2^*)^q$ appearing in the free energy, usual gauge invariance requires $n+m-p-q=0$ and translational invariance requires $n-m-p+q=0$. These two conditions imply that $n=p$ and $m=q$ which leads to the $U(1)\times U(1)$ invariance.  A $U(1)\times U(1)$ symmetry has been examined to discuss possible topological structures in two-band superconductors  \cite{bab02}. Related topological structures have also been discussed in other contexts \cite{sig89,sal85,dim07}.

The  vortices of a $U(1)\times U(1)$ theory can be classified \cite{bab02} by two integers $(n,m)$ which denote a $2n\pi$ phase change in $\psi_1(\br)$ and a $2m\pi$ phase change in $\psi_2(\br)$ as the vortex core is encircled.
Of particular interest here are the $(1,1)$, $(1,0)$, and $(0,1)$ vortices. The $(1,1)$ vortex is the usual Abrikosov vortex and it contains a magnetic flux of $\Phi_0$ (the usual flux quantum).  In the FFLO phase, when $|\psi_1|=|\psi_2|$ (often called the LO phase), the corresponding $(1,0)$ vortex contains a fractional flux $\Phi_0/2$ \cite{bab02}.  We are interested in the appearance of bound pairs of these vortices in the vortex lattice phase. Consequently, we consider generalized Abrikosov vortex lattice states and show the usual FFLO vortex solution is often unstable to a new lattice solution.  In this new solution each of the conventional $(1,1)$ vortices decays into a pair of $(1,0)$ and $(0,1)$ vortices.

We now turn to an analysis valid near the upper critical field. The vortex solutions are eigenstates of the operator $\vD^2=(-i\vnabla-2e\vA)^2$ which has eigenvalues $(2n+1)/l^2$ and $l^2=\Phi_0/(2\pi H)$ and $n=0,1,2,..$ is the LL index. The usual BCS theory predicts a $n=0$ LL solution is the most stable solution, but it has been shown that for FFLO superconductors $n>0$ LL solutions can also be stable \cite{mat07}.  It is well known that the LL exhibit a macroscopic degeneracy.  Abrikosov exploited this degeneracy to construct a vortex lattice solution which we label as $\phi_n(\br)$ \cite{abr59}. We label the unit cell of the vortex lattice by the  lattice vectors $\va=(a,0)$ and $\vb=(b\cos\alpha,b\sin\alpha)$. We take $\br$, $a$, and $b$ to be in units $l$. Then $ab\sin\alpha=2\pi$ gives one flux quantum per unit cell. In this basis, we set
$\vpsi(\br)=[\eta_1\phi_n(\br),\eta_2\tilde{\phi}_n(\br+\vtau)]$ where $\tilde{\phi}_n(\br+\vtau)=e^{-i\tau_yx}\phi_n(\br+\vtau)$. The additional phase factor that appears in $\tilde{\phi}_n$ ensures that both $\psi_1$ and $\psi_2$ lie in the same LL. It appears as a consequence of
applying a translation in a uniform magnetic field. The new feature in this analysis is the appearance of the translation vector $\vtau=(\tau_x,\tau_y)$ that displaces the nodes of the two components $(\psi_1,\psi_2)$. Previous results can be recovered with $\vtau=0$ \cite{hou01,yan04,hou06}. A similar solution has been used for UPt$_3$ \cite{gar94}. Substituting the above solution for $\vpsi(\br)$ yields the free energy density (here we have considered only the $n=0$ LL)
\begin{eqnarray}
f=& \tilde{\alpha_1}|\eta_1|^2+\tilde{\alpha_2}|\eta_2|^2+\tilde{\beta_1}\beta_A(0)|\veta|^4+\nonumber \\&
[(2\tilde{\beta_1}+\tilde{\beta_2})\beta_A(\vtau)-2\tilde{\beta_1}\beta_A(0)]|\eta_1|^2|\eta_2|^2+\nu\gamma_A(0)|\veta|^6\nonumber \\ &+\nu[9\gamma_A(\vtau)-3\gamma_A(0)]|\veta|^2|\eta_1|^2|\eta_2|^2 \label{free2}
\end{eqnarray}
where the coefficients $\tilde{\alpha_1},\tilde{\alpha_2},\tilde{\beta_1},$ and $\tilde{\beta_2}$ do not depend upon  the vortex lattice structure \cite{eq3}. The vortex lattice structure appears entirely in the generalized Abrikosov coefficients $\beta_A(\vtau)=2\pi\int_{u.c.}d^2{\br}|\phi_0(\br)|^2|\phi_0(\br+\vtau)|^2$ and $\gamma_A(\vtau)=(2\pi)^2\int_{u.c.}d^2{\br}|\phi_0(\br)|^4|\phi_0(\br+\vtau)|^2$. Using the approach of Ref.~\cite{yan04} yields
%\begin{equation}
%\beta_A^{(n)}(\vtau)=\sum_{{\bf G}}[L_n(G^2/2)]^2e^{-G^2/2}e^{i{\bf G}\cdot\vtau}
%\end{equation}
\begin{equation}
\beta_A(\vtau)=\sum_{{\bf G}}e^{-G^2/2}e^{i{\bf G}\cdot\vtau}
\end{equation}
and
\begin{equation}
\gamma_A(\vtau)=\sum_{{\bf G},{\bf G}^{\prime}} e^{i\hat{z}\cdot({\bf G}\times{\bf G}^{\prime})/2} e^{i{\bf G}\cdot\vtau}e^{-(G^2+G^{\prime2}+{\bf G}\cdot{\bf G}^{\prime})/2}
\end{equation}
%\begin{eqnarray}
%\gamma_A^{(n)}(\vtau)=&\displaystyle{\sum_{{\bf G},{\bf G}^{\prime}} e^{i\hat{z}\cdot({\bf G}\times{\bf %G}^{\prime})/2} e^{i{\bf G}\cdot\vtau}e^{-(G^2+G^{\prime2}+{\bf G}\cdot{\bf G}^{\prime})/2}} \nonumber \\ &\times  %L_n(G^2/2) L_n(G^{\prime2}/2)L_n(|{\bf G}+{\bf G}^{\prime}|^2/2)
%\end{eqnarray}
where  ${\bf G}=m{\bf g}_1+n{\bf g}_2$ ($n,m$ are any integer), ${\bf g}_1=\sqrt{2\pi\sigma}\hat{x}-\sqrt{2\pi\rho^2/\sigma}\hat{y}$, ${\bf g}_2=\sqrt{2\pi/\sigma}\hat{y}$, and $\rho+i\sigma=e^{i\alpha}b/a$. Below, the ground state lattice structures are numerically found by minimizing Eq.~\ref{free2} with respect $\rho, \sigma,$ and $\vtau$.\\

\noindent {\it Single-$q$ to multiple-$q$ transition in NC superconductors.} Here, $\tilde{\alpha_1}\ne\tilde{\alpha_2}$. When  $\tilde{\alpha_1}<0$ and $\tilde{\alpha_2}>0$, $\eta_1\ne 0$ and $\eta_2=0$, the stable structure is the usual hexagonal vortex lattice.  If $\tilde{\beta_2}<2\tilde{\beta_1}(\beta_A^{(0)}(0)-\beta_A^{0}(\vtau))/\beta_A^{0}(\vtau)$ then a second transition can occur into a state in which both $\eta_1$ and $\eta_2$ are non-zero. This transition has been found within weak-coupling theories of NC superconductors \cite{bar02,dim03,agt07}.  This phase has two possible solutions. The first has $\vtau=0$ and remains a conventional hexagonal lattice. This occurs when $2\tilde{\beta_1}+\tilde{\beta_2}<0$.  The second solution has $\vtau=(\va+\vb)/3$ and occurs for $2\tilde{\beta_1}+\tilde{\beta_2}>0$. To address which of these possibilities occur, we note that weak-coupling  microscopic studies show that the phase diagram contains a line along which $\beta_2=0$  \cite{dim03,agt07}. This implies that the finite $\vtau=(\va+\vb)/3$ phase is the ground state.

\begin{figure}
\epsfxsize=1.8 in \center{\epsfbox{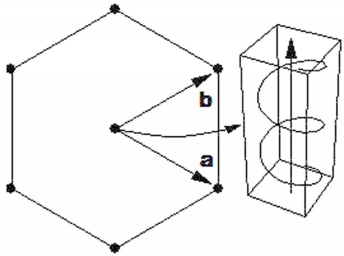}}
\caption{Helical spatial line nodes in the gap for the multiple-$q$ phase of NC superconductors. The centers of the helices form a 2D hexagonal lattice perpendicular to the field.} \label{fig1}
\end{figure}

The spatial nodes of
$\Delta(\vR)=e^{iqz}\eta_1\phi_0(\br)+e^{-iqz}\eta_2\tilde{\phi_0}
(\br+\vtau)$ are given by
$|\eta_1\phi_0(\br)|=|\eta_2\phi_0(\br+\vtau)|$ and
$\cos[qz+(\theta_1-\theta_2)/2]=0$ where $\theta_1=\theta_1(\br)$
is the phase of $\phi_0(\br)$ and $\theta_2(\br)$ is the phase of
$\tilde{\phi_0} (\br+\vtau)$. For small $\eta_2$,  these zeroes
lie on small circles surrounding each of the zeroes of $\psi_1$.
Around these circles, the phase $\theta_1(\br)=\phi$ since we are
encircling a vortex core of $\psi_1$ (here, $\phi$ is the polar
angle of the circle) and $\theta_2\approx cnst$ since we are far
away from the zeroes of $\psi_2(\br)$. Consequently, the nodes of
$\Delta(\vR)$ are given by $qz=\phi/2+n\pi+c$ where $c$ is a
constant and $n$ is any integer. This describes the equation of a
helix spiralling about the $z$ direction. This is depicted in
Fig.~1. As $\eta_2$ grows, the pitch of the helix grows larger. It
is possible for two adjacent helices to merge for large enough
$\eta_2$. This results in a crisscrossing lattice of line nodes
like that discussed below in the context of the FFLO case. This
analysis reveals that the $(n,m)=(1,1)$ Abrikosov vortices have
each separated into a pair of $(1,0)$ and $(0,1)$ vortices. The
$(1,0)$ vortices appear where $\psi_1(\br)=0$ and the $(0,1)$
vortices appear where $\psi_2(\br)=0$.

\noindent {\it Second order transition into the FFLO phase.} Here,
$\alpha_1=\alpha_2$, $\nu=0$, and there is one second order
transition from the normal state into the FFLO state. There are
three possible solutions for this phase. The first has $\eta_2=0$
and $\eta_1\ne0$, this is the FF (or single-$q$) state with a
conventional hexagonal lattice. This phase is stable when
$(2\tilde{\beta}_1+\tilde{\beta}_2)\beta_A(\vtau)-2\tilde{\beta}_1\beta_A(0)>0$
for all $\vtau$. One of the other two solutions are stable if
$(2\tilde{\beta}_1+\tilde{\beta}_2)\beta_A(\vtau)-2\tilde{\beta}_1\beta_A(0)<0$
for any $\vtau$. The second solution corresponds to
$|\eta_1|=|\eta_2|$ with $\vtau=0$ and is the LO (or multiple-$q$)
phase  with a conventional hexagonal lattice. This state requires
$2\tilde{\beta}_1+\tilde{\beta}_2<0$ to be stable. The final state
corresponds to $|\eta_1|=|\eta_2|$ with a rectangular unit cell
for which $b/a=\sqrt{3}$ and $\vtau=(a,b)/2$ when
$\tilde{\beta}_2=0$. More generally (including first order FFLO
transitions) we find the same lattice but with $b/a\ne\sqrt{3}$.
These solutions are stable for
$2\tilde{\beta}_1+\tilde{\beta}_2>0$. To understand which of these
states may be stable within microscopic theories, note that the
calculations of Ref.~\cite{hou01} imply that there is a line in
the phase diagram along which $\tilde{\beta_2}=0$ in the
weak-coupling theory of a clean $s$-wave superconductor with
vortices. Near this line, $\vtau=(a,b)/2$ gives the stable phase.
Whenever the FF phase is close in energy to the LO phase (that is
$|\tilde{\beta}_2|<<\tilde{\beta}_1$), then the LO vortex phase
with $\vtau=(a,b)/2$ is the stable vortex phase since
$\beta_A(\vtau)\le\beta_A(0)$ for any $\vtau\ne0$. It appears that
this is generic for weak-coupling theories where varying gap
symmetry, impurities, and vortices lead to a variety of different
phase diagrams containing both the FF and LO phases
\cite{buz97,hou01,hou06,agt01}.
% comments perhaps should mention that \beta(\vtau)\le \beta(0)
% also mention that \tilde{\beta}_1>0

% bewteen differences seen in recent NMR measurements on CeCoIn$_5$ \cite{kak05,mit06}. The $\vtau=(a,b)/2$ may also %provide an explanation for unexplained thermal conductivity measurements in CeCoIn$_5$ \cite{cap04}.

We now focus on the LO phase with $\vtau=(a,b)/2$. This phase can
be understood as having conventional $(1,1)$ Abrikosov vortices
that have each separated into a pair of $(1,0)$ and $(0,1)$
vortices. As discussed previously the $(1,0)$ and $(0,1)$ vortices
in this LO phase can be interpreted as containing flux $\Phi_0/2$.
To understand if this may manifest itself experimentally,  we have
performed an Abrikosov analysis \cite{abr59} on Eq.~\ref{free1} to
determine the field distribution $h_s(\br)\hat{z}$ due to
screening currents to lowest order in the gap function. This
results in $h_s(\br)\propto |\psi_1(\br)|^2+|\psi_2(\br)|^2$.
Consequently, when $\tilde{\beta}_2=0$, $h_s(\br)$  has a
hexagonal symmetry even though the nodes  of $\psi_1(\br)$ and
$\psi_2(\br)$ separately form a rectangular lattice with
$b/a=\sqrt{3}$. A measurement of the hexagonal unit cell lattice
vector will yield a flux per unit cell that is  $\Phi_0/2$ (this
generalizes to non-hexagonal unit cell geometries). This can be
seen through SANS measurements by observing the Bragg peaks of the
vortex lattice with neutrons that have momenta perpendicular to
the applied field. We emphasize that our solution is valid at
$H_{c2}$ and at lower fields it is possible that the $\Phi_0/2$
vortices are more tightly bound (e.g. $\vtau\ne0$ but
$|\vtau|<|\va+\vb|/2$).

\begin{figure}
\epsfxsize=2.0 in \center{\epsfbox{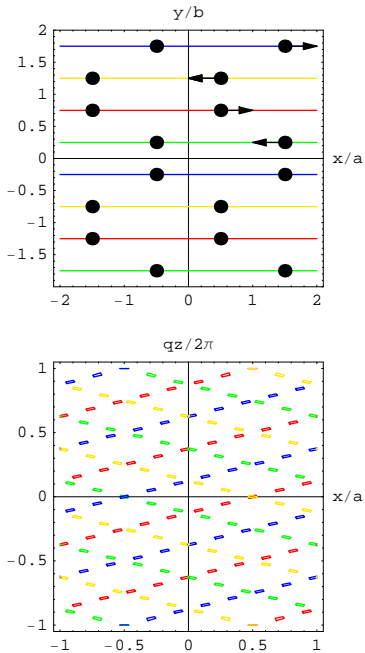}} \caption{(Color
online) Crisscrossing lattice of nodal lines in the FFLO vortex
phase with $\vtau=(a,b)/2$. The dark circles in the top figure
shows the nodes perpendicular to the applied magnetic field for
$z=0$. As $z$ is changed slightly, these nodes move as illustrated
by the arrows in the upper right of this figure. The lower figure
shows a cross section as seen from the $y$ direction. The
different colors lines correspond to the nodal lines with
different $y$ positions.}   \label{fig2}
\end{figure}

 Here we give the positions of the line nodes for $\vtau=(a,b)/2$ and $b/a>3$.
 In the $x,y$ plane the point zeros lie along the lines
$y_1=-3b/4$, $y_2=-b/4$, $y_3=b/4$, and $y_4=3b/4$ (the unit cell
has doubled along the y-direction). The $x,z$ coordinates
(measured in units $a,\pi/q$ respectively) for these four lines
are given by $z_1=n+1/2-x_1/2$, $z_2=n+1/2+x_2/2$, $z_3=n-x_3/2$,
and $z_4=n+x_4/2$. This results in a lattice of crisscrossing
nodal lines as viewed from a direction normal to the $y$-axis
(Fig.~\ref{fig2}).

For the FFLO phase in CeCoIn$_5$, the $\vtau=(a,b)/2$ solutions may help to understand some experiments \cite{mat07}. In particular, measurements in the FFLO vortex phase find that the thermal conductivity parallel to the applied field is greater than that perpendicular to the applied field \cite{cig05}. This is not expected for a gap function with spatial plane nodes perpendicular to the field (which occurs if $\vtau=0$). However, it can be qualitatively understood if  $\vtau=(a,b)/2$. Note that magnetic order in the FFLO nodal planes has been proposed \cite{you07} and this may account for the thermal conductivity results when $\vtau=0$.

In conclusion, we have argued that the vortex lattice phases in FFLO and NC superconductors contain gap functions with spatial line nodes that form a variety of three dimensional spatial configurations. These configurations include a lattices of helices in NC superconductors and a crisscrossing lattice of nodal lines in FFLO superconductors. These structures stem from the break up  of conventional vortices into pairs of fractional vortices. SANS studies of the  magnetic field distribution can provide evidence for these structures.

We thank Cigdem Capan, Manfred Sigrist, Ilya Vekhter, Anton Voronstov, and Kun Yang for useful discussions.
This work was supported by the National Science Foundation grant No. DMR-0381665.

\end{document}